\documentclass[12pt,a4paper,titlepage,oneside]{article}
\usepackage{cite, rotate}
\usepackage{amsmath}
\usepackage[dvips]{graphics}
\usepackage[dvips]{graphicx}
\usepackage[bf]{subfigure}
\begin{document}

\begin{center}
\noindent \textbf{\large Quaternionic Representation of Snub 24-Cell and its
  Dual Polytope Derived From $E_{8}$ Root System}  \\
\end{center}



{\bf Mehmet Koca$^1$, Mudhahir Al-Ajmi$^1$, Nazife Ozdes Koca$^1$} \\
$^1$Department of Physics, College of Science, Sultan Qaboos  \\
University, P. O. Box 36, Al-Khoud, 123 Muscat, Sultanate of Oman \\

{\bf Email}: kocam@squ.edu.om, mudhahir@squ.edu.om, nazife@squ.edu.om

\section*{Abstract}
\noindent Vertices of the 4-dimensional semi-regular polytope,
\textit{snub 24-cell} and its symmetry group $W(D_{4} ):C_{3} $ of
order 576 are represented in terms of quaternions with unit norm. It
follows from the icosian representation of \textbf{$E_{8} $} root
system. A simple method is employed to construct the \textbf{$E_{8} $}
root system in terms of icosians which decomposes into two copies of
the quaternionic root system of the Coxeter group $W(H_{4} )$, while
one set is the elements of the binary icosahedral group the other set
is a scaled copy of the first. The quaternionic root system of $H_{4}
$ splits as the vertices of 24-cell and the \textit{snub 24-cell}
under the symmetry group of the \textit{snub 24-cell} which is one of
the maximal subgroups of the group \textbf{$W(H_{4} )$} as well as
$W(F_{4} )$. It is noted that the group is isomorphic to the\textbf{
}semi-direct product of the Weyl group of $D_{4}$ with the cyclic
group of order 3 denoted by $W(D_{4} ):C_{3} $, the Coxeter notation
for which is $[3,4,3^{+}]$.  We analyze the vertex structure of the
\textit{snub 24-cell} and decompose the orbits of \textbf{$W(H_{4} )$}
under the orbits of $W(D_{4} ):C_{3} $. The cell structure of the snub
24-cell has been explicitly analyzed with quaternions by using the
subgroups of the group $W(D_{4} ):C_{3} $. In particular, it has been
shown that the dual polytopes 600-cell with 120 vertices and 120-cell
with 600 vertices decompose as 120=24+96 and 600=24+96+192+288
respectively under the group $W(D_{4} ):C_{3} $. The dual polytope of
the \textit{ snub 24-cell} is explicitly constructed.  Decompositions
of the Archimedean $W(H_{4} )$ polytopes under the symmetry of the
group $W(D_{4} ):C_{3} $ are given in the appendix.

\section{Introduction}

\noindent The non-crystallographic Coxeter group $W(H_{4} )$ has some
relevance to the quasicrystallography~[1,2,3]. It is one of the
maximal subgroups of $W(E_{8})$~[4], the Weyl group of the exceptional
Lie group $E_{8} $ which seems to be playing an important role in high
energy physics~[5]. The symmetries $A_{4}, B_{4}, F_{4}$ occur in high
energy physics in model building either as a gauge symmetry like
$SU(5)$~[6] or as a little group $SO(9)$~[7] of M-theory. The
exceptional Lie group $F_{4}$ is also in the domain of interest of
high energy physicists~[7,8]. The Coxeter-Weyl groups $W(H_4), W(A_4),
W(B_4)$ and $W(F_4)$ describe the symmetries of the regular 4D
polytopes. In particular, the Coxeter group $W(H_{4} )$ arises as the
symmetry group of the polytope 600-cell, $\{ 3,3,5\}$~[9], vertices of
which can be represented by 120 quaternions of the binary icosahedral
group~[10, 11]. The dual polytope 120-cell, $\{ 5,3,3\} $with 600
vertices, can be constructed from 600-cell in terms of
quaternions~[11]. The symmetries of the the 4D polytopes can be nicely
described using the finite subgroups of quaternions~[12].

\noindent 

\noindent In this paper we study the symmetry group of the
semi-regular 4D polytope \textit{snub 24-cell} and construct its 96
vertices in terms of quaternions. It is a semi-regular polytope with
96 vertices, 432 edges, 480 faces of equilateral triangles and 144
cells of two types which was first discovered by Gosset~[13]. We
explicitly show how these vertices form 120 tetrahedral and 24
icosahedral cells which constitute the \textit{snub 24-cell}. We
organize the paper as follows. In Section 2 we construct the root
system of \textbf{$E_{8} $} using \textbf{ }two sets of quaternionic
representations of the roots as well as the weights of three
8-dimensional representations of $D_{4}$~[14]. This construction leads
to two copies of quaternionic representations of the vertices of
600-cell.  In Section 3 we introduce the quaternionic root system of
$H_{4}$ decomposed in terms of its conjugacy classes where classes
correspond to the orbits of the Coxeter group $W(H_{3})$ in which the
vertices of the \textit{icosidodecahedron }are represented by
imaginary quaternions~[15]. We review the cell structures of the
600-cell and the 120-cell using the conjugacy classes of the binary
icosahedral group. In Section 4 we construct the maximal subgroup
$W(D_{4} ):C_{3}$ of $W(H_{4} )$ using the quaternionic vertices of
24-cell corresponding to the quaternionic elements of the binary
tetrahedral group and work out explicitly the decomposition of the
600-cell as well as 120-cell under $W(D_{4} ):C_{3}$. Section 5 is
devoted to the explicit study of the cell structures of the
\textit{snub 24 cell}. We construct the dual polytope of \textit{snub
  24 cell} in Section~6 in terms of quaternions. Remarks and
discussions are given in the conclusion. In the appendix we give
decompositions of the regular and semi-regular orbits of $W(H_4)$
under the group $W(D_4):C_3$.

\noindent 

\section{Construction of the root system of $E_{8}$ in terms of icosians}

\noindent \textbf{}

\noindent Let $q=q_{0} +q_{i} e_{i} $, ($i=1,{\rm \; }2,{\rm \; }3)$
be a real quaternion with its conjugate defined by $\bar{q}=q_{0}
-q_{i} e_{i}$ where the quaternionic imaginary units satisfy the
relations

\begin{equation} \label{GrindEQ__1_} 
e_{i} e_{j} =-\delta _{ij} +\varepsilon _{ijk} e_{k} , (i,j,k=1,{\rm \; }2,{\rm \; }3).                                                        
\end{equation} 
Here $\delta _{ij}$ and $\varepsilon _{ijk} $ are the Kronecker and
Levi-Civita symbols respectively and summation over the repeated
indices is implicit. Quaternions generate the four dimensional
Euclidean space where the quaternionic scalar product can be defined
as

\begin{equation} \label{GrindEQ__2_} 
(p,q)=\frac{1}{2} (\bar{p}q+\bar{q}p).                                                                     
\end{equation} 
The group of unit quaternions is isomorphic to $SU(2)$ which is a
double cover of the proper rotation group $SO(3)$.  The imaginary
quaternionic units $e_{i}$ can be related to the Pauli matrices
$\sigma _{i}$ by $e_{i} =-i\sigma _{i}$ and the unit quaternion is
represented by a $2\times 2$ unit matrix. The roots of $D_{4}$ can be
obtained from the Coxeter-Dynkin diagram (Figure~1) where the scaled
simple roots are denoted by quaternions of unit norm

\begin{equation} \label{GrindEQ__3_} 
\alpha _{1} =e_{1} ,{\rm \; }\alpha _{2} =\frac{1}{2} (1-e_{1} -e_{2} -e_{3} ),{\rm \; }\alpha _{3} =e_{2} ,{\rm \; }\alpha _{4} =e_{3}.  
\end{equation}

\begin{figure}[h]
\begin{center}
  \includegraphics[height=3cm]{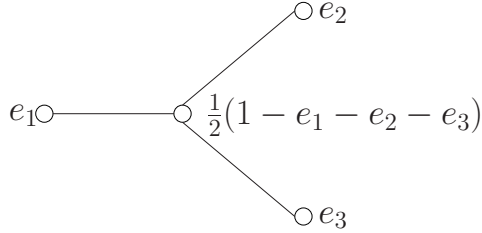}
\end{center}
\caption{Coxeter-Dynkin diagram of $D_{4} $with quaternionic simple
  roots.}
\end{figure}

\noindent This will lead to the following orbits of weights of $D_{4}$
under the Weyl group $W(D_4)$:

\begin{equation} \label{GrindEQ__4_} 
  \begin{array}{l} {O(1000):V_{1} =\{ \frac{1}{2} (\pm 1\pm e_{1} ),\frac{1}{2} (\pm e_{2} \pm e_{3} )\} } \\ {O(0010):V_{2} =\{ \frac{1}{2} (\pm 1\pm e_{2} ),\frac{1}{2} (\pm e_{3} \pm e_{1} )\} } \\ {O(0001):V_{3} =\{ \frac{1}{2} (\pm 1\pm e_{3} ),\frac{1}{2} (\pm e_{1} \pm e_{2} )\} } \\ {O(0100):T=\{\pm 1, \pm e_{1} ,\pm e_{2} ,\pm e_{3} ,\frac{1}{2} (\pm 1\pm e_{1} \pm e_{2} \pm e_{3} )\} }. \\
\end{array}
\end{equation} 

Let us also define the set of quaternions ${T'=\{ \sqrt{2} V_{1}
  \oplus \sqrt{2} V_{2} \oplus \sqrt{2} V_{3} \} }$, dual of $T$.

\noindent We will adopt the Lie algebraic notation $\Lambda=(a_{1}
a_{2} ...a_{n} )$ for the highest weight~[16] for a Lie group $G$ of
rank \textit{n} but the notation $O(a_{1} a_{2} ...a_{n} )=W(G)(a_{1}
a_{2} ...a_{n})$ stands for the orbit deduced from the highest weight.
The first three sets of quaternions in (\ref{GrindEQ__4_}) are known
as the weights of three 8-dimensional vector, spinor, and antispinor
representations of $SO(\ref{GrindEQ__8_})$ and the last 24 quaternions
are the non-zero roots of the same Lie algebra.  Actually the set of
quaternions in (\ref{GrindEQ__4_}) constitute the non-zero roots of
$F_{4} $. When the set of quaternions of $V_{i} (i=1,2,3)$ are
multiplied by $\sqrt{2}$ and taken together with the set \textit{T},
they constitute the elements of the binary octahedral group of order
48. The set \textit{T} alone represents the binary tetrahedral group
of order 24 and also they are the\textbf{ }vertices of the 24-cell
whose symmetry is the group $W(F_{4} )$ of order 1152.

\noindent The \textbf{$E_{8} $} root system can be constructed in
terms of two sets of quaternionic root systems of $F_{4}$ as
follows~[4]:

\begin{equation} \label{GrindEQ__5_} 
(T,0)\oplus (0,T)\oplus (V_{1} ,V_{3} )\oplus (V_{2} ,V_{1} )\oplus (V_{3} ,V_{2} )  
\end{equation}

\noindent where the ordered pair $(A,B)$ means $A+\sigma B$ and
$\sigma =\frac{1-\sqrt{5} }{2}$ with the golden ratio $\tau
=\frac{1+\sqrt{5} }{2}$ they satisfying the relations $\tau \sigma
=-1$, $\tau +\sigma =1$, $\tau ^{2} =\tau +1$ and $\sigma ^{2} =\sigma
+1$ and $\oplus$ represents the union of the sets. The 240 quaternions
in (\ref{GrindEQ__5_}) can be written as the union of $I\oplus \sigma
I$~[4] where \textit{I} stands for the set of 120 quaternionic
elements of the binary icosahedral group which also constitutes the
root system of $H_{4}$ representing the vertices of 600-cell. Any two
quaternions $p,q$ from (\ref{GrindEQ__5_}) satisfy the scalar product
$(p,q)=x+\sigma y$ where $x,y=0,\pm \frac{1}{2} ,\pm 1$. If one
defines the Euclidean scalar product$(p,q)_{E} =x$~[17] then the set
of quaternions in (\ref{GrindEQ__5_}) represents the roots of
\textbf{$E_{8} $}. This proves that the roots of \textbf{$E_{8} $}
which correspond to the vertices of the \textbf{$E_{8} $-} Gosset's
polytope decompose as two copies of 600-cells in 4 dimensional
Euclidean space, one is a scaled copy of the other. If one adopts the
Euclidean scalar product then $\sigma I$ and \textit{I} lie in two
orthogonal 4D spaces.

\noindent 

\noindent

\section{Quaternions and $W(H_{4} )$}

\noindent It is well known that the set of icosians \textit{I}
constitute the roots of $H_{4} $ and can be generated from the Coxeter
diagram where the simple roots are represented by unit
quaternions~[11]. In our analysis the subgroup $W(H_{3} )$ plays a
crucial role. Therefore we want to display the elements of \textit{I}
as the orbits of $W(H_{3} )$. They are tabulated in Table 1 in
terms of its conjugacy classes of the binary icosahedral group, in
other words as the orbits of $W(H_{3} )$. The conjugacy classes
represent the vertices of four icosahedra, two dodecahedra and one
icosidodecahedron as well as two single points $\pm 1$.

\begin{table}[h]
  \caption[r]{\normalsize{Conjugacy classes of the binary icosahedral group $I$ represented by quaternions}}

\begin{tabular}{lll}
  \hline
  Conjugacy classes and &&   The sets of the conjugacy classes denoted \\
  orders of elements    &&   by the number of elements \\
  \hline
  1  & $1$ &   \\  
  2  & $-1$ &   \\  
  10 &  $12_+:$ & $\frac{1}{2} (\tau \pm e_{1} \pm \sigma e_{3} ),\frac{1}{2} (\tau \pm e_{2} \pm \sigma e_{1} ),$  \\
     &  &  $\frac{1}{2} (\tau \pm e_{3} \pm \sigma e_{2} )$ \\
  5  &   $12_-:$ & $\frac{1}{2} (-\tau \pm e_{1} \pm \sigma e_{3} ),\frac{1}{2} (-\tau \pm e_{2} \pm \sigma e_{1} ),$ \\
     &  &  $\frac{1}{2} (-\tau \pm e_{3} \pm \sigma e_{2}$ ) \\
  10 &   $12_+^\prime:$ & $\frac{1}{2} (\sigma \pm e_{1} \pm \tau e_{2} ),\frac{1}{2} (\sigma \pm e_{2} \pm \tau e_{3} ),$ \\
     &  &  $\frac{1}{2} (\sigma \pm e_{3} \pm \tau e_{1} )$ \\ 
  5  &   $12_-^\prime:$ & $\frac{1}{2} (-\sigma \pm e_{1} \pm \tau e_{2} ),\frac{1}{2} (-\sigma \pm e_{2} \pm \tau e_{3} ),$ \\ 
     &  &          $\frac{1}{2} (-\sigma \pm e_{3} \pm \tau e_{1} )$  \\
  6  &   $20_+:$ & $\frac{1}{2} (1\pm e_{1} \pm e_{2} \pm e_{3} ),\frac{1}{2} (1\pm \tau e_{1} \pm \sigma e_{2} ),$  \\  

     &   & $\frac{1}{2} (1\pm \tau e_{2} \pm \sigma e_{3} ),\frac{1}{2} (1\pm \tau e_{3} \pm \sigma e_{1} )$ \\  
  3  & $20_-:$ & $\frac{1}{2} (-1\pm e_{1} \pm e_{2} \pm e_{3} ),\frac{1}{2} (-1\pm \tau e_{1} \pm \sigma e_{2} ),$  \\ 
     &  & $\frac{1}{2} (-1\pm \tau e_{2} \pm \sigma e_{3} ),\frac{1}{2} (-1\pm \tau e_{3} \pm \sigma e_{1} )$ \\  
  4  &  $30:$ & $\pm e_{1} ,\pm e_{2} ,\pm e_{3} ,\frac{1}{2} (\pm \sigma e_{1} \pm \tau e_{2} \pm e_{3} ),$ \\ 
     &  & $\frac{1}{2} (\pm \sigma e_{2} \pm \tau e_{3} \pm e_{1} ),\frac{1}{2} (\pm \sigma e_{3} \pm \tau e_{1} \pm e_{2} )$ \\  
       
  \hline
\end{tabular}

\end{table}

\noindent Denote by $p,q\in I$ any two unit quaternions. One can
define a transformation on an arbitrary quaternion \textit{r} by

\begin{equation} \label{GrindEQ__6_} 
[p,q]:r\to r'=prq,~[p,q]^{*} :r\to r''=p\bar{r}q.
\end{equation} 

The set of all these transformations constitute the Coxeter group

\noindent 

\begin{equation} \label{GrindEQ__7_} 
W(H_{4} )=\{ [p,q]\oplus [p,q]^{*};~p,q\in I\}  
\end{equation}

\noindent of order 14,400~[11]. It is clear that the set \textit{I}
itself is invariant under the group $W(H_{4} )$. The binary tetrahedral
group \textit{T} is one of the maximal subgroups of \textit{I}. Let
\textit{p} be an arbitrary element of \textit{I} which satisfies $p^{5}
=\pm 1$. Note that we have the relations

\noindent 

\begin{equation} \label{GrindEQ__8_} 
\bar{p}=\pm p^{4},~\bar{p}^{2} =\pm p^{3},~\bar{p}^{3} =\pm p^{2},~\bar{p}^{4} =\pm p.                         
\end{equation}

\noindent Then one can write the set of elements of the binary
icosahedral group, in other words, the vertices of the 600-cell
as~[10, 11]

\noindent                           

\begin{equation} \label{GrindEQ__9_} 
I=\sum _{j=0}^{4}\oplus p^{j}  T=T\oplus \sum _{j=1}^{4}\oplus p^{j}  
T=T\oplus \sum _{j=1}^{4}\oplus  T\bar{p}^{j} .                                
\end{equation} 
This decomposition is true when $p^{j} $ is replaced by any conjugate
$p^{j} _{c} =qp^{j} \bar{q}$ with $q\in I$.

\noindent \textbf{}

\section{Embedding of the group $W(D_{4} ):C_{3} $ in the Coxeter group $W(H_{4} )$}

\noindent The five conjugate groups of the group $T$ can be
represented by the sets of elements $p^{i} T\bar{p}^{i} $ (
$i=0,1,2,3,4$). The sets $p^{i} T$ are the five copies of the 24-cell
in the 600-cell.  It is then natural to expect that the Coxeter group
\textbf{$W(H_{4} )$} contains the group with the set of 576 elements
$\{ [T,T]\oplus [T,T]^{*} \} $ as a maximal subgroup~[18]. One can
prove that it is the extension of the Coxeter-Weyl group $W(D_{4} )$
of order 192 by the cyclic group $C_{3} $ which permutes, in the
cyclic order, three outer simple roots $e_{1} ,e_{2} ,e_{3} $ of
$D_{4} $ in Figure 1. In reference~[18] it was shown that the group is
isomorphic to the semi-direct product of these two groups, namely, the
group $ W(D_{4} ):C_{3} =\{ [T,T]\oplus [T,T]^{*} \}$. The group
$W(D_{4} ):C_{3} $ is also a maximal subgroup of the group $W(F_{4}
)\approx W(D_{4} ):S_{3} $ of order 1152 which is the symmetry group
of the 24-cell~[8]. The 25 conjugate groups of the group $W(D_{4}
):C_{3} =\{ [T,T]\oplus [T,T]^{*} \} $ in the group \textbf{$W(H_{4}
  )$} can be represented by the groups

\begin{eqnarray} \label{GrindEQ__10_} 
(W(D_{4} ):C_{3} )^{(i,j)} & = & \{ [p^{i} T\bar{p}^{i} ,p^{j} 
T\bar{p}^{j} ]\oplus [p^{i} T\bar{p}^{j} ,p^{i} T\bar{p}^{j} ]^{*} \} ; \nonumber \\
(i,j=0,1,2,3,4)
\end{eqnarray}

with \textit{p} satisfying $p^{5} =\pm 1$. Without loss of generality,
we will work with the group $(W(D_{4} ):C_{3} )^{(0,0)} =W(D_{4}
):C_{3} =\{ [T,T]\oplus [T,T]^{*} \}$.

\noindent Let us denote by 

\begin{equation} \label{GrindEQ__11_} 
S=I-T=\sum _{j=1}^{4}\oplus p^{j}  T=\sum _{j=1}^{4}\oplus  Tp^{j}  
\end{equation} 
the set of 96 quaternions representing the vertices of the
\textit{snub 24-cell}. The above discussion indicates that the\textit{
  snub 24-cell} can be embedded in the 600-cell five different ways.
Of course when two copies of 24-cell $T\oplus \sigma T$ is removed
from the root system of \textbf{$E_{8} $}, in other words from
the\textbf{ }Gosset's polytope, what remain are two copies of the
\textit{snub 24-cell}

\noindent 

\begin{eqnarray} \label{GrindEQ__12_} 
\{ O(1000)+\sigma O(0001)\} &\oplus& \{O(0001)+\sigma O(0010)\} \nonumber \\
  &\oplus& \{ O(0010)+\sigma O(1000)\} = S\oplus \sigma S.
\end{eqnarray} 

We will work with the set\textit{ S} because the other is just a
scaled copy of\textit{ S}. It is clear that the set \textit{T}
representing the 24-cell is invariant under the group $W(D_{4} ):C_{3}
$ which acts on the set \textit{S }as follows:

\begin{eqnarray} \label{GrindEQ__13_} 
[T,T]:S\Rightarrow  \sum _{j=1}^{4}\oplus Tp^{j} TT 
& = & \sum _{j=1}^{4}\oplus p_{c} ^{j}  T=\sum _{j=1}^{4}\oplus  Tp_{c} ^{j}  
\end{eqnarray} 
   and  
\begin{equation} \label{GrindEQ__14_} 
[T,T]^{*} :S\Rightarrow \sum _{j=1}^{4}\oplus T\bar{T}\bar{p}^{j}  
T=\sum _{j=1}^{4}\oplus  T\bar{p}_{c} ^{j} =\sum _{j=1}^{4}\oplus p_{c} ^{j}  
T=\sum _{j=1}^{4}\oplus  Tp_{c} ^{j}              
\end{equation} 
                                              
where $p_c$ is an arbitrary conjugate of $p$.
\noindent It implies that the set \textit{S} is invariant under the
group $W(D_{4} ):C_{3} $. In the Coxeter's notation this group is
denoted by $[3,4,3^{+} ]$~[19].

\noindent It was shown in the references~[10,11] that the set of 600
vertices of the 120-cell can be written as the set of quaternions
$J=\sum _{i,j=0}^{4}\oplus p^{i} T'\bar{q}^{j} =\sum _{i=0}^{4}\oplus
p^{i} \bar{p}^{\dag i} I,~(i,j=0,1,2,3,4)$ where the quaternions
\textit{p} and \textit{q} satisfy the relations $p^{5} =q^{5} =\pm 1$
and $p=\bar{c}{\rm \; }\bar{q}^{\dag } c$. Here $p^{\dag } $ is
obtained from \textit{p} by replacing $\sigma \leftrightarrow \tau $
and \textit{c} is an arbitrary element of $T'$. It is not difficult to
show that under the group $ W(D_{4} ):C_{3}$ it decomposes as

\begin{eqnarray} \label{GrindEQ__15_} 
J=\sum _{i,j=0}^{4}\oplus p^{i}
  \bar{p}^{\dag j} T'=T'&+&\sum _{i=1}^{4}\oplus p^{i} \bar{p}^{\dag i} 
   T' +\sum _{i=1}^{4}(\oplus p^{i} T' \oplus \bar{p}^{\dag i}
   T') \nonumber \\
&+&\sum _{i\ne j=1}^{4}\oplus p^{i} \bar{p}^{\dag j} T'.
\end{eqnarray} 
These are the orbits of $W(D_{4} ):C_{3} $ of sizes 24, 96, 192, and
288. Let us denote the orbits of sizes 96, 192 and 288 in 120-cell
by $S'=\sum _{i=1}^{4}\oplus p^{i} \bar{p}^{\dag i} T' ,M=\sum
_{i=1}^{4}(\oplus p^{i} T' \oplus \bar{p}^{\dag i} T'),N=\sum
_{i\ne j=1}^{4}\oplus p^{i} \bar{p}^{\dag j} T'$ respectively. We shall
discuss later that the orbits $T,T',S'$ of sizes 24, 24, 96
respectively are important for the \textit{dual snub 24-cell}.
\\
\noindent We now discuss the cell structure of the \textit{snub
  24-cell}~[20]. In reference~[11] we have given a detailed analysis
of the cell structures of the 4D polytopes 600-cell and 120-cell which
are orbits of the Coxeter group $W(H_{4} )$. There, we have proved
that the conjugacy classes in those two examples denoted by $12_{+} $
can be decomposed as 20 sets of 3-quaternion system, each set is
representing an equilateral triangle. The vertices of all equilateral
triangles are equidistant from the unit quaternion 1, implying that
one obtains a structure of 20 tetrahedra meeting at the vertex 1. The
centers of the 20 tetrahedra constitute the vertices of a
dodecahedron. The symmetry of a tetrahedron is the group $W(A_{3})
\approx S_4$ of order 24.  Therefore the number of tetrahedra
constituting the 600-cell is the index of $W(A_{3} )$ in $W(H_{4} )$
that is 600.  The set $12_{+} $, representing an icosahedron in three
dimensions, is indeed an orbit of the icosahedral subgroup $W(H_{3} )$
whose index in $W(H_{4} )$ is 120. Those 20 sets of quaternions in
$12_{+} $ represent the faces of the icosahedron, five of which meet
at one vertex. This shows that each quaternion of $12_{+} $ belongs to
5 of those 20 sets. The decomposition of quaternions in Table 1 is
made with respect to a subgroup $W(H_{3} )$ which leaves the unit
quaternion~1 invariant. The group $W(H_3)$ can be represented as the
subset of elements of $W(H_{4} )$ by

\noindent              

\begin{equation} \label{GrindEQ__16_} 
W(H_{3} )=\{ [I,\bar{I}]\oplus [I,\bar{I}]^{*} \} .                                         
\end{equation}

\noindent In fact, one of the maximal subgroup of $W(H_{4} )$ is
$W(H_{3} )\times C_{2}$ of order 240 where $C_{2} $, in our notation,
is generated by the group element $[1,-1]$. In the Coxeter group
$W(H_{4} )$ there are 60 conjugates of the group $W(H_{3} )\times
C_{2} $, each one is leaving one pair of elements $\pm q$ invariant.
The conjugates of the group $W(H_{3} ) \times C_{2}$ can be
represented compactly by the set of group elements

\noindent 

\begin{equation} \label{GrindEQ__17_} 
\{ W(H_{3} )\times C_{2} \} ^{q} =\{ [I,\pm \bar{q}\bar{I}q]\oplus [I,\pm q\bar{I}q]^{*} \},~q \in I.                           
\end{equation}

\noindent The orbits of $W(H_{3} )^{q}$ in \textit{I} can be written
as $\pm q,q(12_{\pm } )\equiv 12_{\pm } (q),~q(12'_{\pm } )\equiv
12'_{\pm } (q),q(20_{\pm } )\equiv 20_{\pm } (q),q(30)\equiv 30(q)$.
The set of quaternions in the conjugacy classes are multiplied by the
quaternion~$q$ on the left or on the right. Let $t\in T\subset I$,
then we can form the set $t(12_{+} )$ in 24 different ways. Each set
$t(12_{+} )$ together with one \textit{$t\in T$ }represents 20
tetrahedra. Therefore, with the set of quaternions of \textit{T}, each
is sitting at one vertex; one obtains $24 \times 20=480$ tetrahedra.
Actually the centers of these 480 tetrahedra lie on the 480=192+288
vertices of the \textit{120-cell}. Removing 480 tetrahedra from
\textit{I} results in removal of the sets M and N from the set
\textit{J}.  We know that the set $t(12_{+} )$ does not involve any
quaternion from \textit{T}. If that were the case then the scalar
product of \textit{t} with this element would yield to $\frac{\tau
}{2}$.  But this is impossible for any two elements of the set
\textit{T}.  When the vertices of \textit{T} are removed from the
600-cell the remaining 120 tetrahedral cells belong to the\textit{
  snub 24-cell}. It is clear then that when the quaternion
\textit{$t\in T$} is removed what is left in the void is the
icosahedron represented by the vertices $t(12_{+} )$. Therefore
instead of 480 tetrahedra there has been created 24 icosahedra, the
vertices of which are the sets $t(12_{+} )$ with \textit{t }taking 24
values in the set \textit{T.}  Adding this to the remaining 120
tetrahedral cells then the number of cells of the\textit{ snub
  24-cell} will be 144. Below we will work out the detailed cell
structure of the polytope of concern.

\noindent 

\section{Detailed analysis of the cell structure of the \textit{snub 24-cell}}\textit{}

\noindent The conjugacy class $12_{+}(1)$ can be written as the products of three elements; denote the elements by 

\begin{eqnarray} \label{GrindEQ__18_} 
p=\frac{1}{2} (\tau +e_{1} +\sigma e_{3} ),t_{1} =\frac{1}{2} (1+e_{1} -e_{2} -e_{3} ), \nonumber \\
t_{2} =\frac{1}{2} (1+e_{1} +e_{2} -e_{3} ).
\end{eqnarray} 

The set of three elements where $t_{1} ,t_{2} $ and $t_{1} t_{2}
=e_{1} $ are elements of \textit{T}. Then we can represent the set of
elements of $12_{+}(1) $ by

\noindent 

\begin{eqnarray} \label{GrindEQ__19_} 
12_{+} (1)&=&\{ p,p^{2} (\bar{t}_{2} \bar{t}_{1} ),~\bar{p}^{2} t_{1} ,~\bar{p}^{2} t_{2},~p\bar{t}_{2},~p^{2} \bar{t}_{2},~p^{2} \bar{t}_{1}, \nonumber  \\
& &  ~\bar{p}^{2} t_{1} t_{2},~p\bar{t}_{1},~\bar{p},~\bar{p}t_{1},~\bar{p}t_{2} \}.                   
\end{eqnarray}

As we noted earlier this set of elements represents an icosahedron.
Now multiply this set either from left or right by $t_{1} $ and $t_{2}
$ to obtain the sets

\begin{eqnarray} \label{GrindEQ__20_} 
12_{+} (t_{1} )&=&\{ pt_{1}
  ,~p^{2} \bar{t}_{2} ,~\bar{p}^{2} t_{1} ^{2} ,~\bar{p}^{2} t_{2} t_{1}
  ,~p\bar{t}_{2} t_{1} ,~p^{2} \bar{t}_{2} t_{1} ,~p^{2} ,~ \nonumber \\
  & &  \bar{p}^{2} t_{1} t_{2} t_{1} ,~p,~\bar{p}t_{1} ,~\bar{p}t_{1} ^{2} ,~\bar{p}t_{2}t_{1} \}
\end{eqnarray} 
\begin{eqnarray} \label{GrindEQ__21_} 
12_{+} (t_{2} )&=&\{ pt_{2} ,~p^{2}
  (\bar{t}_{2} \bar{t}_{1} t_{2} ),~\bar{p}^{2} t_{1} t_{2}
  ,~\bar{p}^{2} t_{2} ^{2} ,~p,~p^{2} ,  \nonumber \\
  & & p^{2} \bar{t}_{1} t_{2} ,~\bar{p}^{2} t_{1} t_{2} ^{2} ,~p\bar{t}_{1} t_{2} ,~\bar{p}t_{2},~\bar{p}t_{1} t_{2} ,~\bar{p}t_{2} ^{2} \}.
\end{eqnarray} 

\noindent Each set represents an icosahedron. Here the sets in (20)
and (21) are left invariant respectively by the conjugate groups of
$W(H_{3} )$, namely by, $W(H_{3} )^{t_{1} } =\{ [I,\bar{t}_{1}
\bar{I}t_{1} ]\oplus [I,t_{1} \bar{I}t_{1} ]^{*} \} $ and $W(H_{3}
)^{t_{2} } =\{ [I,\bar{t}_{2} \bar{I}t_{2} ]\oplus [I,t_{2}
\bar{I}t_{2} ]^{*} \} $ respectively. The centers of these three
icosahedra are represented, up to a scale factor, by three quaternions
$1,t_{1}$ and $t_{2} $ respectively. The crucial thing in (19-21) is
that all three icosahedra have one common vertex represented by the
quaternion \textit{p}. This proves that three icosahedra meet at one
vertex \textit{p}. Now we show that there are exactly five tetrahedra
meeting at the vertex \textit{p}. When we multiply the set of elements
in (19) by \textit{p} we obtain the set

\noindent 

\begin{eqnarray} \label{GrindEQ__22_} 
12_{+} (p)&=&\{ p^{2} ,~p^{3} (\bar{t}_{2} \bar{t}_{1} ),~\bar{p}t_{1} ,~\bar{p}t_{2} ,~p^{2} \bar{t}_{2} ,~p^{3} \bar{t}_{2} ,\\ \nonumber
& &~p^{3} \bar{t}_{1} ,~\bar{p}t_{1} t_{2} ,~p^{2} \bar{t}_{1} ,~1,~t_{1} ,~t_{2} \}.
\end{eqnarray} 
We have already explained that the set of elements in (22) form the
set of 20 tetrahedra, all connected to the vertex represented by
\textit{p}. We also note that 15 of them have vertices involving the
quaternions $1,t_{1} ,t_{2}$ which are elements of 24-cell \textit{T}.
The remaining set of five tetrahedra involves the vertices belonging
only to the set \textit{S}. The above arguments indicate that, at the
given vertex \textit{p,} the \textit{snub 24-cell} has five
tetrahedral and three icosahedral cells. This is true for all elements
of the set \textit{S}. Therefore the \textit{snub 24-cell} has
$\frac{96\times 3}{12} =24$ icosahedral cells and $\frac{96\times
  5}{4} =120$ tetrahedral cells as explained before with a total
number of 144 cells.

\noindent 

\noindent One can look at the problem from the symmetry point of view
of the cells. One of the maximal subgroups of $\{ W(H_{3} )\}^{q}$ in
(17) is obtained when \textit{I} is restricted to \textit{T} which can
be written in the form

\begin{equation} \label{GrindEQ__23_} 
\{A_{4} \times C_{2}\}^q =\{ [T,\bar{q}\bar{T}q]\oplus [T,q\bar{T}q]^{*} \}.                     
\end{equation} 
Here $A_{4} $ stands for the even permutations of the four letters, a
group of order 12, and together with the cyclic group $C_{2}$, it is a
group of order 24 which permutes the vertices of icosahedron leaving
its center represented by the quaternion $q\in T$ invariant. The group
in (23) can be embedded in the symmetry group $W(D_{4} ):C_{3} $ of
the \textit{snub 24-cell} in 24 different ways, in each case leaving
one quaternion $q\in T$ invariant. This shows that the number of
icosahedral cells of the \textit{snub 24-cell} is 24 and the
centers of these icosahedra are represented, up to a scale factor, by
the set of quaternions belonging to the set \textit{T}. The group in
(23) is not the full symmetry of an icosahedron but just a subgroup of
it because the symmetry group of the \textit{snub 24-cell} does not
involve the whole symmetry group of the icosahedron. As we noted
before in 600-cell, \textit{I}, the number of vertices closest to a
given vertex is 12 while this number in the \textit{snub 24-cell} that
is in the set \textit{S} is 9 as shown in (22). The 9 vertices nearest
to the quaternion \textit{p} in (22) represent the vertices of five
tetrahedra as well as the five nearest vertices of three icosahedra.
Let us denote them by

\[q_{1} =\frac{1}{2} (-\sigma +\tau e_{1} -e_{3} ), q_{2} =\frac{1}{2} (\tau -\sigma e_{1} -e_{2} ),q_{3} =\frac{1}{2} (\tau -\sigma e_{1} +e_{2} ),\] 
\begin{equation} \label{GrindEQ__24_} 
q_{4} =\frac{1}{2} (\tau +\sigma e_{2} -e_{3} ),q_{5} =\frac{1}{2} (\tau -\sigma e_{2} -e_{3} ),q_{6} =\frac{1}{2} (1-\sigma e_{1} -\tau e_{3} ),
\end{equation} 
\[q_{7} =\frac{1}{2} (\tau +e_{1} -\sigma e_{3} ),q_{8} =\frac{1}{2} (1+\tau e_{1} -\sigma e_{2} ), q_{9} =\frac{1}{2} (1+\tau e_{1} +\sigma e_{2} ).\] 

\noindent The vertices of five tetrahedra $P(i)~(i=1,2,3,4,5)$ meeting
at the point $q_{10} \equiv p=\frac{1}{2} (\tau +e_{1} +\sigma e_{3}
)$ can be obtained from (24) and their corresponding centers
$c_i~(i=1,...,5)$ can be written up to a scale factor as follows:

\begin{equation}\label{GrindEQ__25_}
\begin{array}{l}
P(1)=\{q_{10} ,q_{7} ,q_{8} ,q_{9}\};~c_{1} =\frac{1}{\sqrt{2} } (1+e_{1} ), \\
P(2) =\{q_{10} ,q_{4} ,q_{5} ,q_{6}\};~c_{2}=\frac{1}{2\sqrt{2} } ((\tau -\sigma )-\sigma e_{1} -\tau e_{3}),  \\
P(3) =\{q_{10} ,q_{3} ,q_{7} ,q_{8}\};~c_{3} =\frac{1}{2\sqrt{2} } (\tau -\sigma +\tau e_{1} -\sigma e_{2}),  \\
P(4) =\{q_{10} ,q_{1} ,q_{8} ,q_{9}\};~c_{4} =\frac{1}{2\sqrt{2} } (\tau +(\tau -\sigma )e_{1} +\sigma e_{3}),  \\
P(5) =\{q_{10} ,q_{2} ,q_{7} ,q_{9}\};~c_{5} =\frac{1}{2\sqrt{2} } (\tau -\sigma +\tau e_{1} +\sigma e_{2} ).
\end{array}
\end{equation}
Now one can check that there exist 15 equilateral triangles having
\textit{p} as a common vertex. Then the number of faces of the
\textit{snub 24-cell} is $\frac{15\times 96}{3} =480$. Similarly the
number of edges are $\frac{9\times 96}{2} =432$.

\noindent The nine vertices $q_{i}~(i=1,2,...,9)$ closest to the
quaternion \textit{p} represent the vertices of the \textit{vertex
  figure }of the \textit{snub 24-cell} as we will discuss later. It is
interesting to note that the center of the tetrahedron $P(1)$ belongs
to the set $T'$ which is invariant under the symmetry group of the
\textit{snub 24-cell}, $W(D_{4} ):C_{3} $.  Action of the group
$W(D_{4} ):C_{3} $ on the tetrahedron $P(1)$ will generate 24
tetrahedra whose centers, up to a scale factor, lie on the vertices of
the 24-cell, $T'$. One can show that the centers of the tetrahedra
$P(i)~(i=2,3,4,5)$ belong to the set of 96 vertices of the set $S'$.
The subgroup of the group $W(D_{4} ):C_{3} $ preserving the
tetrahedron $P(1)$ can be written in the form

\noindent 

\begin{equation} \label{GrindEQ__26_} 
S_{4} =\{ [T,\bar{c}_{1} \bar{T}c_{1} ]\oplus [T,c_{1} \bar{T}c_{1} ]^{*} \} ,c_{1} =\frac{1}{\sqrt{2} } (1+e_{1} )\in T'.             
\end{equation} 
This is the tetrahedral subgroup of $W(D_{4} ):C_{3} $ which is
isomorphic to the symmetric group $S_{4} $ of four letters. It is the
group which permutes the vertices of the tetrahedron $P(1)$ while
fixing its center $c_{1} =\frac{1}{\sqrt{2} } (1+e_{1} )$. The
conjugates of the group $S_{4} $ represented by (26) are the 24
different subgroups each fixing one element of $T'$. This is another
proof that there exist 24 tetrahedra of type $P(1)$ whose centers lie
on the orbit $T'$.  It is also interesting to note that the centers of
the 24 icosahedral cells lie on the vertices of the other 24-cell
represented by the set \textit{T} which can be obtained from $T'$ by
rotation around some axis.

\noindent 

\noindent The subgroup which preserves one of the remaining 96
tetrahedra is a group isomorphic to the symmetric group $S_{3} $. This
is not a surprise because the index of $S_{3} $ in the group $W(D_{4}
):C_{3} $ is 96 corresponding to the 96 tetrahedra.

\noindent Let us construct the subgroup of $W(D_{4} ):C_{3} $ which
fixes the vertex $q_{10} \equiv p=\frac{1}{2} (\tau +e_{1} +\sigma
e_{3} )=\frac{1}{2} [(1+e_{1} )-\sigma (1-e_{3} )]=\frac{1}{2} [\tau
(1+e_{1} )+\sigma (e_{1} +e_{3} )]$ implying that \textit{p} can be
written as a linear combination of two elements of the set $T'$. This
is evident because we have already noted this property in (12) which
followed from (5). Another interesting relation similar to (12) can be
written in the form

\begin{eqnarray} \label{GrindEQ__27_} 
\{ \tau O(1000)+\sigma O(0010)\} &\oplus& \{ \tau O(0010)+\sigma O(0001)\} \nonumber \\ 
  &\oplus& \{ \tau O(0001)+\sigma O(1000)\} =S\oplus S'.
\end{eqnarray} 

\noindent This shows that the elements of two sets \textit{S} and $S'$
can be written, up to a scale factor, in the form of $\tau a+\sigma b$
where \textit{a} and \textit{b }are elements of the set $T'$. We have
already shown that the subgroup of the group $W(D_{4} ):C_{3} $
leaving one element of $T'$ invariant can be written as $S_{4} =\{
[T,\bar{a}\bar{T}a]\oplus [T,a\bar{T}a]^{*} \}$. A subgroup of this
group which also fixes the element \textit{b} leaves the quaternion
$\tau a+\sigma b$ invariant. To find the generators we apply the above
group elements on \textit{b}, namely,

\noindent 

\begin{equation} \label{GrindEQ__28_} 
[t,\bar{a}\bar{t}a]:b\to
  tb\bar{a}\bar{t}a=b;~[t,a\bar{t}a]^{*} :b\to t\bar{b}a\bar{t}a=b .
\end{equation} 
where $t \in T$.

\noindent From the first relation it follows that
$b\bar{a}\bar{t}=\bar{t}b\bar{a}$ implying that $t=b\bar{a}$ is one of
the solutions besides the quaternion 1. The second relation will lead
to $\bar{b}a\bar{t}=\bar{t}b\bar{a}$. For the case of \textit{p} above
one can choose $a=\frac{1}{\sqrt{2} } (1+e_{1} ),b=\frac{1}{\sqrt{2} }
(e_{1} +e_{3} )$ which leads to the element $b\bar{a}\equiv s_{1}
=\frac{1}{2} (1+e_{1} -e_{2} +e_{3} ),\;\bar{b}a\equiv s_{2}
=\frac{1}{2} (1-e_{1} -e_{2} -e_{3} )$ of $T$ and the group generator
would be

\begin{equation} \label{GrindEQ__29_} 
[s_{1} ,s_{2} ].                                    
\end{equation} 
The second relation implies that the group element of \textit{T} must be
$e_{2} $ and the group generator reads

\begin{equation} \label{GrindEQ__30_} 
[e_{2} ,-e_{2} ]^{*}.
\end{equation} 
One can prove that the generators in (29) and in (30) generate a group
of order 6 isomorphic to the symmetric group $S_{3} $. It is also
interesting to note that the same group fixes the quaternions $c_{1} $
and $c_{2} $ corresponding to the centers of two tetrahedra $P(1)$ and
$P(2)$ which implies that these two tetrahedra are left invariant by
the group fixing the quaternion \textit{p}. One can show that the
tetrahedra $P(3)$, $P(4)$, $P(5)$ are permuted by the group $S_{3}$.
Similarly the three icosahedra $12_+(1),12_+(t_{1}),12_+(t_{2} )$
given in (19-21) are permuted among themselves.  This group is also
the symmetry of the vertex figure of the \textit{snub 24-cell} as we
will discuss it now.

\noindent 

\noindent The vertex figure of any convex polytope is the convex solid
formed by the nearest vertices to it. In our case they are the
quaternions $q_{i}~(i=1,2,...,9)$. Since all these quaternions have
the same scalar product, $\tau /2$, with the quaternion \textit{p} they
all lie in the same hyperplane orthogonal to \textit{p}.  One should
then express them in an orthogonal basis involving \textit{p}.  The
new set of basis can be obtained by multiplying the quaternionic units
$1,e_{1} ,e_{2} ,e_{3} $ by \textit{p} on the right or left and define
the new basis as follows

\begin{equation} \label{GrindEQ__31_} 
\begin{array}{l} {p_{0} =p=\frac{1}{2} (\tau +e_{1} +\sigma e_{3} ),p_{1} =e_{1} p=\frac{1}{2} (-1+\tau e_{1} -\sigma e_{2} ),} \\ {p_{2} =e_{2} p=\frac{1}{2} (\sigma e_{1} +\tau e_{2} -e_{3} ),p_{3} =e_{3} p=\frac{1}{2} (-\sigma +e_{2} +\tau e_{3} ).} \end{array} 
\end{equation} 

\noindent When 9 quaternions $q_{i} ,{\rm \; }i=1,2,...,9$ are
expressed in terms of the new basis vectors and the first component
multiplying $p_{0} $ is removed then the nine quaternion with
remaining components would read

\begin{equation} \label{GrindEQ__32_} 
\begin{array}{l} {(\pm 1,0,\sigma ),(1,0,-\sigma ),(\sigma ,\pm 1,0),(0,\sigma ,1),(-\sigma ,-1,0),(0,-\sigma ,\pm 1)}. \end{array} 
\end{equation} 

An overall scale factor $\frac{1}{2}$ is omitted. These vertices
represent the \textit{tridiminished icosahedron}, a Johnson's solid,
$J_{63} $~[21] as shown in Figure 2(a) and Figure 2(b). If three more
vertices $(-1,0,-\sigma ),(-\sigma ,1,0),(0,\sigma ,-1)$ are added to
(32) we would have the vertices of an icosahedron as shown in Figure
2(c).

\begin{figure}[h]
\begin{center}
  \centering 
      \subfigure[]{\includegraphics[height=4cm]{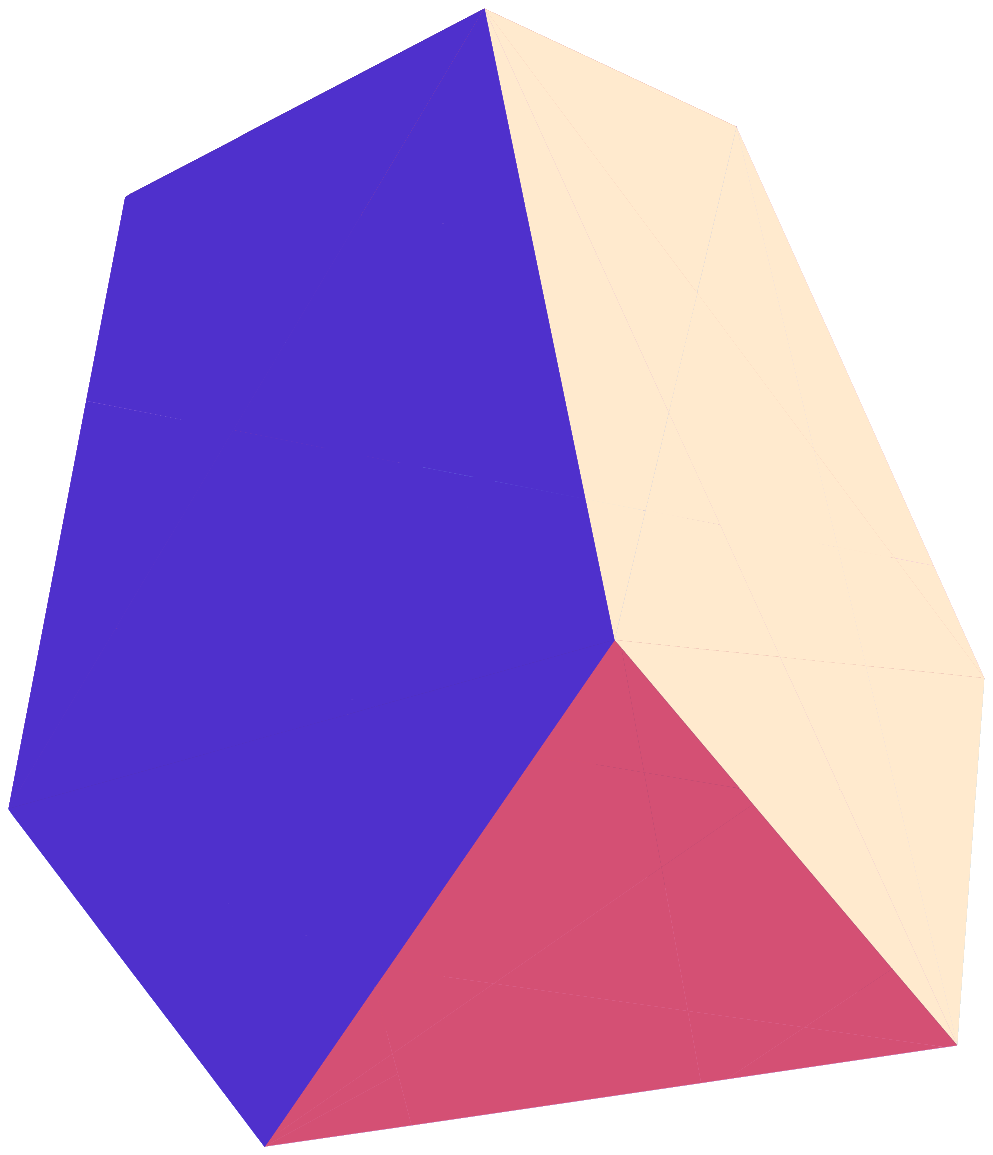}}
      \subfigure[]{\includegraphics[height=4cm]{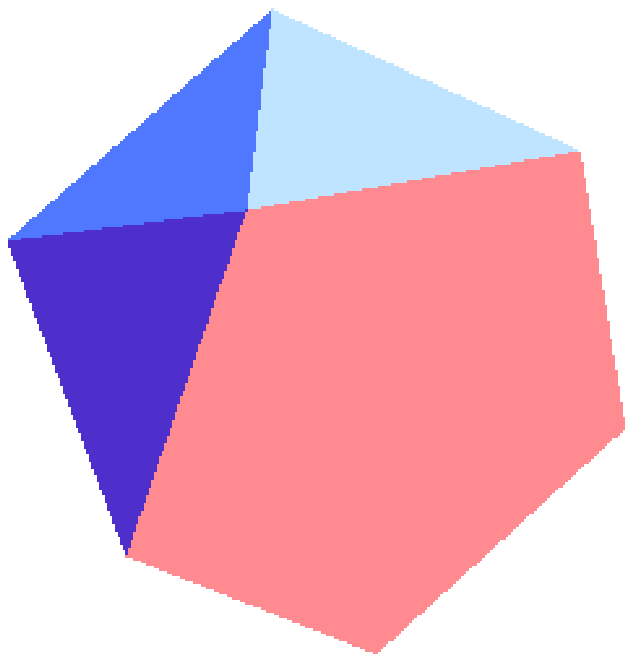}}\\
      \subfigure[]{\includegraphics[height=4cm]{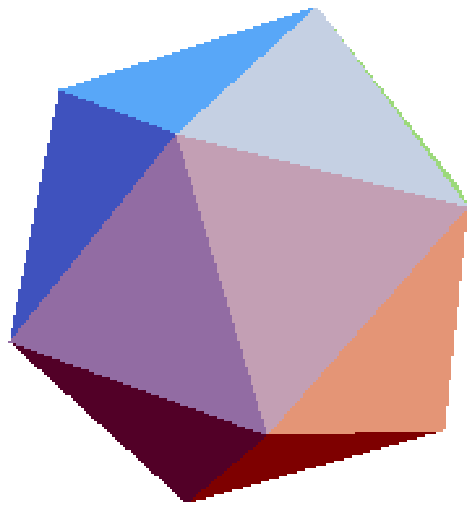}}
\end{center}
\caption{(a) Tridiminished icosahedron. (b) Tridiminished 
icosahedron (another view). (c) Icosahedron}
\end{figure}

\noindent A net of the tridiminished icosahedron is depicted in Figure
3 where the vertices identified with those nine quaternions.

\begin{figure}[h]
\begin{center}
  \centering 
  \includegraphics[height=4.5cm]{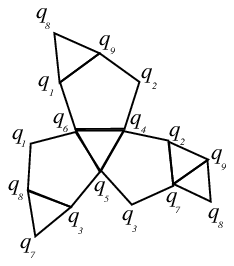}
\end{center}
\caption{The net of tridiminished icosahedron where the vertices are identified with nine quaternions}
\end{figure}

\noindent Removing those three vertices from an icosahedron reduces
the symmetry of the icosahedron of order 120 to the symmetry of order
6. The tridiminished icosahedron has three pentagonal and five
triangular faces. Its symmetry is the symmetric group $S_{3} $,
generated by the generators given by (29-30), which permutes the
vertices of each set among themselves:

\begin{equation} \label{GrindEQ__33_} 
(q_{1} ,q_{2} ,q_{3} ),(q_{4} ,q_{5} ,q_{6} ),(q_{7} ,q_{8} ,q_{9}). 
\end{equation}

\noindent The symmetry axis goes through the centers of two opposite
triangular faces represented by $(q_{4} ,q_{5} ,q_{6} ),(q_{7} ,q_{8}
,q_{9} )$. Three pentagonal faces and the remaining three triangular
faces are permuted by $S_{3} $.

\section{Dual of the \textit{snub 24-cell}}

\noindent Dual polytope of a given regular or semi-regular polytope is
constructed by taking the centers of its cells as the vertices of the
dual polytope. Since \textit{snub-24 cell }has 144 cells, the
\textit{dual snub-24 cell} will have 144 vertices. In order that the
dual cell has the same symmetry group of its original polytope, its
vertices should lie in the hyperplanes orthogonal to the vertices of
the original polytope. We have studied the Catalan solids which are
duals of the Archimedean solids in a different paper in the context of
quaternions~[22]. The vertices of the dual cell of a semi-regular
polytope have, in general, different lengths lying on the concentric
3-spheres $S^{3} $ with different radii. Here, in our example, we
should determine hyperplane orthogonal to the vertex, say again,
\textit{p}.  The centers of five tetrahedra, up to a scale factor, are
already determined by the unit quaternions $c_{i}~(i=1,2,3,4,5)$ given
in (25).  We recall that the centers of those three icosahedra lie on
the unit vectors $1,{\rm \; }t_{1} ,{\rm \; }t_{2} $ up to a scale
factor.  When we multiply the latter three unit quaternions by
$\frac{\tau }{\sqrt{2} }$, then all eight quaternions $\frac{\tau
}{\sqrt{2} } 1,\frac{\tau }{\sqrt{2} } t_{1} ,\frac{\tau }{\sqrt{2} }
t_{2} $,$~c_{i}~(i=1,2,3,4,5)$ lie in the same hyperplane determined
by the equation $\tau q_{0} +q_{1} +\sigma q_{3} =\frac{\tau ^{2}
}{\sqrt{2} } $ which is orthogonal to the vertex \textit{p}. Now we
use the basis defined in (31) to express above eight quaternions as
follows:

\begin{equation} \label{GrindEQ__34_} 
\begin{array}{l}
\frac{\tau }{\sqrt{2} } 1=\frac{1}{2\sqrt{2} } (\tau ^{2} p_{0}
-\tau p_{1} +p_{3} ), \\
\frac{\tau }{\sqrt{2} } t_{1} =\frac{1}{2\sqrt{2}
} (\tau ^{2} p_{0} -p_{2} -\tau p_{3} ), \\
\frac{\tau }{\sqrt{2} } t_{2}
=\frac{1}{2\sqrt{2} } (\tau ^{2} p_{0} +p_{1} +\tau p_{2} ), \\
c_{1} =\frac{1}{2\sqrt{2} } (\tau ^{2} p_{0} -\sigma p_{1} +\sigma
p_{2} -\sigma p_{3} ), \\
c_{2} =\frac{1}{2\sqrt{2} } (\tau ^{2} p_{0}
+\sigma p_{1} -\sigma p_{2} +\sigma p_{3} ),  \\
c_{3} =\frac{1}{2\sqrt{2}} (\tau ^{2} p_{0} +\sigma ^{2} p_{1} +p_{3} ), \\
c_{4} = \frac{1}{2\sqrt{2} } (\tau ^{2} p_{0} +p_{1} -\sigma ^{2} p_{2} ), \\
c_{5} = \frac{1}{2\sqrt{2} } (\tau ^{2} p_{0} -p_{2} +\sigma ^{2} p_{3} ).
\end{array} 
\end{equation}
After removing the components along $p=p_{0} $ and omitting the
factor $\frac{1}{2\sqrt{2} } $ one writes the 8 vertices in the above
order as follows:

\begin{eqnarray*}
\begin{array}{l}
  (-\tau ,0,1),(0,-1,-\tau ),(1,\tau ,0),
  (-\sigma ,\sigma ,-\sigma ),  \\
  (\sigma ,-\sigma ,\sigma ), (\sigma ^{2} ,0,1),
  (1,-\sigma ^{2} ,0),(0,-1,\sigma ^{2} ).
\end{array}
\end{eqnarray*}

\noindent These are the vertices of one cell of the dual polytope of
the \textit{snub 24- cell} shown in Figure 4(a) and Figure 4(b).

\begin{figure}[h]
\begin{center}
  \centering 
      \subfigure[]{\includegraphics[height=4cm]{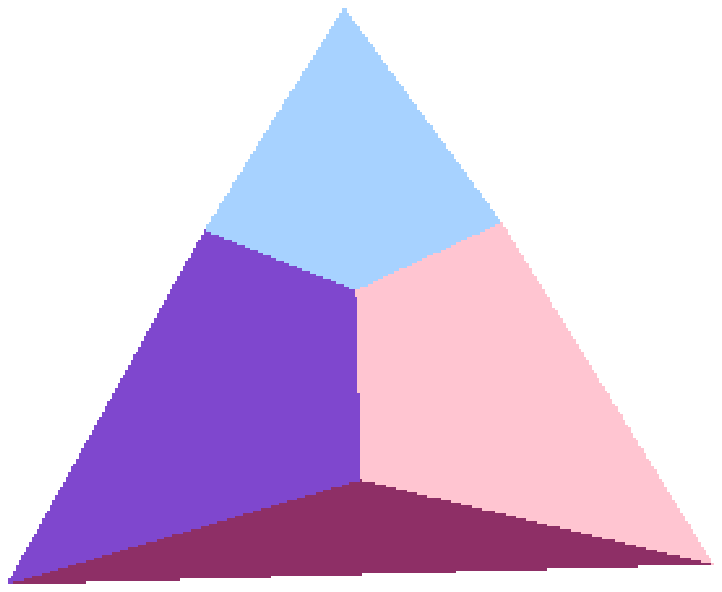}}
      \subfigure[]{\includegraphics[height=4cm]{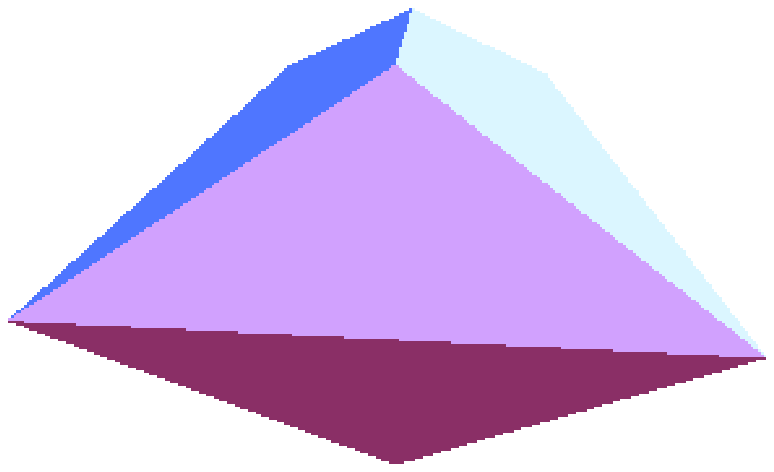}}
\end{center}
\caption{(a) Cell of the \textit{dual} \textit{snub 24-cell}. (b) Cell of the \textit{dual} \textit{snub 24-cell} (another view)}
\end{figure}

\noindent It is clear from the Figures 4(a)-4(b) that the cell has three
kite faces with edge lengths $\frac{1}{\sqrt{2} } $ and $\frac{\sigma
  ^{2} }{\sqrt{2} } $ with the shorter diagonal of length $\frac{\sigma
}{\sqrt{2} } $ and six isosceles triangles with edge lengths
$\frac{1}{\sqrt{2} } $ and $\frac{\tau }{\sqrt{2} } $ as shown in the
Figure 5(a) and Figure 5(b).

\begin{figure}[h]
\begin{center}
  \centering 
      \subfigure[]{\includegraphics[height=3cm]{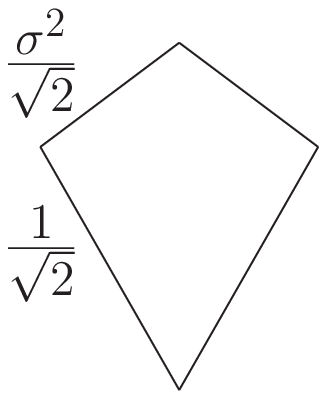}}
      \subfigure[]{\includegraphics[height=2.5cm]{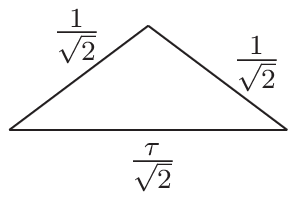}}
\end{center}
\caption{(a) Kite face of the cell of \textit{dual snub 24-cell}.  (b)
  Isosceles triangular face of the cell of \textit{dual snub 24-cell}}
\end{figure}

\noindent The symmetric group $S_{3}$, fixing the vertex \textit{p}
also fixes the vertices $c_{1} $and $c_{2} $ and permutes each set of
three vertices $\frac{\tau }{\sqrt{2} } 1,\frac{\tau }{\sqrt{2} }
t_{1} ,\frac{\tau }{\sqrt{2} } t_{2} $ and $(c_{3} ,c_{4} ,c_{5} )$.
We have 96 cells of the same type constituting the \textit{dual}
\textit{snub-24 cell}. It is clear that the \textit{dual snub 24-
  cell} is cell transitive for the group $W(D_{4} ):C_{3} $ is
transitive on the vertices of the \textit{snub 24- cell}. The vertices
of three kites are given as follows:

\noindent 

\begin{equation} \label{GrindEQ__35_} 
(c_{1} ,c_{3} ,c_{5} ,\frac{\tau}{\sqrt{2}} 1);~(c_{1} ,c_{4} ,c_{5} ,\frac{\tau}{\sqrt{2}} t_{1} );~(c_{1} ,c_{3} ,c_{4} ,\frac{\tau}{\sqrt{2}} t_{2} )  .                                          
\end{equation}

\noindent The vertices of the six isosceles triangles are represented
by sets of three quaternions:

\begin{equation*}
(c_{2} ,\frac{\tau }{\sqrt{2} } 1,\frac{\tau }{\sqrt{2} }
t_{1} );~(c_{2} ,\frac{\tau }{\sqrt{2} } t_{1} ,\frac{\tau }{\sqrt{2} }
t_{2} );~(c_{2} ,\frac{\tau }{\sqrt{2} } t_{2} ,\frac{\tau }{\sqrt{2} }
1);
\end{equation*}

\begin{equation}
(c_{5} ,\frac{\tau }{\sqrt{2} } 1,\frac{\tau }{\sqrt{2} }
t_{1} );~(c_{4} ,\frac{\tau }{\sqrt{2} } t_{1} ,\frac{\tau }{\sqrt{2} }
t_{2} );~(c_{3} ,\frac{\tau }{\sqrt{2} } t_{2} ,\frac{\tau }{\sqrt{2} }
1).
\end{equation}

\noindent The $S_{4}$ symmetry, fixing the vertex $c_{1}
=\frac{1}{\sqrt{2} } (1+e_{1} )$, when applied on the vertex $c_{2} $,
will generate four vertices surrounding the vertex $c_{1} \in T'$.
Since $c_{2} $ is fixed by the $S_{3} $ subgroup, one generator,
say $[1,e_{1} ]^{*} $ of $S_{4} $, not belonging to $S_{3} $, would
suffice to determine the four quaternions when applied on $c_{2} $.
They are given by the set of quaternions:

\begin{equation} \label{GrindEQ__37_} 
\begin{array}{l}
c_{2} =\frac{1}{\sqrt{2} } (\sqrt{5} -\sigma e_{1} -\tau e_{3} ),~
c'_{2} =\frac{1}{\sqrt{2} } (-\sigma +\sqrt{5} e_{1} +\tau e_{2} ), \\ 
c''_{2} =\frac{1}{\sqrt{2} } (\sqrt{5} -\sigma e_{1} +\tau e_{3} ),~ 
c'''_{2} =\frac{1}{\sqrt{2} } (-\sigma +\sqrt{5} e_{1} -\tau e_{2}). 
\end{array}
\end{equation}

\noindent This means there are four cells connected to the vertex
$c_{1} \in T'$ while only one cell is connected to any vertex of the
set $S'$. These four vertices form a tetrahedron when $c_{1} $ is taken
to be the origin. Similarly the group fixing $c_{1} $ will generate six
vertices belonging to the set of quaternions \textit{T} which can be
written as

\begin{equation} \label{GrindEQ__38_} 
1,t_{1} ,t_{2} ,e_{1} ,s_{1} ,s_{2}  .                                                                                           
\end{equation} 
They form an octahedron around the vertex $c_{1} $. When we apply the
group $S_{4} $ on three vertices $c_{3} ,c_{4} ,c_{5} $ we will have
one more vertex $c_{6} =\frac{1}{2\sqrt{2} } (\tau +\sqrt{5} e_{1}
-\sigma e_{3} )$. These four vertices also form a tetrahedron around
the vertex $c_{1} $. One should note that the three sets of vertices
$(c_{3} ,c_{4} ,c_{5} ,c_{6} )$ and those given in (37) and (38) lie
in three parallel hyperplanes orthogonal to $c_{1} $. The cell of the
\textit{dual snub 24-cell} which is rotated by the group generator
$[1,e_{1} ]^{*}$ can be displayed by the set of quaternions as
follows:

\begin{equation}
\begin{tabular}{ccccccc}
\(
c_1\) & \(\to\) &  \(c_1\) &  \(\to\) &  \(c_1\) &  \(\to\) &  \(c_1\) \\ \\
\(\left(\begin{array}{c} {c_{3} } \\ {c_{4} } \\ {c_{5} } \end{array}\right)\) &  \(\to\) &  \(\left(\begin{array}{c} {c_{6} } \\ {c_{3} } \\ {c_{4} } \end{array}\right)\) &  
\(\to\) &  \(\left(\begin{array}{c} {c_{5} } \\ {c_{6} } \\ {c_{3} } \end{array}\right)\) &  \(\to\) &  \(\left(\begin{array}{c} {c_{4} } \\ {c_{5} } \\ {c_{6} } \end{array}\right) \) \\ \\
\(\frac{\tau}{\sqrt{2}} \left(\begin{array}{c} {1 } \\ {t_1 } \\ {t_2 } \end{array}\right)\) &  \(\to\) &  \(\frac{\tau}{\sqrt{2}}\left(\begin{array}{c} {e_1 } \\ {t_2 } \\ {s_2 } \end{array}\right)\) &  
\(\to\) &  \(\frac{\tau}{\sqrt{2}}\left(\begin{array}{c} {1 } \\ {s_2 } \\ {s_1 } \end{array}\right)\) &  \(\to\) &  \(\frac{\tau}{\sqrt{2}}\left(\begin{array}{c} {e_1 } \\ {s_1 } \\ {t_1 } \end{array}\right) \) \\ \\
\(c_2\) & \(\to\) &  \(c_2^{\prime}\) &  \(\to\) &  \(c_2^{\prime\prime}\) &  \(\to\) &  \(c_2^{\prime\prime\prime}\) \\
\end{tabular}
\end{equation}

One can count the number of edges and faces of the \textit{dual snub
  24-cell}. The number of faces is 432 made of 144 kites and 288
isosceles triangles while the number of edges is 480.

\noindent 

\section{Conclusion }

\noindent The beauty of \textit{snub 24-cell} lies in the fact that it
contains both types of cells, tetrahedral cells as well as icosahedral
cells. While the first one is relevant to the crystallography, the
second one describes the quasi crystallography. Such structures are
rare. To obtain the \textit{snub-24 cell} from the root system of
$E_{8} $ we followed a chain of symmetry $W(D_{4} ):C_{3} \subset
W(H_{4} )\subset W(E_{8} )$, each being a maximal subgroup in the
larger group.  This is a new approach which has not been discussed in
the literature although it has been known that the removal of a
24-cell from 600-cell leads to the \textit{snub 24- cell}. Another
novel approach in our discussion is that we use quaternions to
describe the vertices as well as the group elements acting on the
vertices. One does not need any computer calculation to obtain the
vertices and the symmetries of the polytope of concern and its dual
except in plotting of three dimensional cells. To the best of our
knowledge the structure of the \textit{dual snub 24-cell} has not been
discussed elsewhere. The decomposition of the Platonic as well as the
Archimedean $H_{4} $ orbits under the symmetry group $W(D_{4} ):C_{3}
$ has been given in the Appendix.

\begin{appendix}
\section*{Appendix: Decomposition of the Platonic and Archimedean orbits of $W(H_{4} )$ under the group $W(D_{4} ):C_{3} $}

Orbit 1:\\ 600 =
192+
96+
288+
24. \\
Orbit 2:\\ 1200 =
144+
576+
288+
2(96). \\
Orbit 3:\\ 720 =
2(288)+
144. \\
Orbit 4:\\ 120 =
24+
96. \\
Orbit 5:\\ 3600 =
5(576)+
4(288)+
144. \\
Orbit 6:\\ 2400 =
2(576)+
2(96)+
3(288)+
192. \\
Orbit 7:\\ 3600 =
144+
4(576)+
4(288). \\
Orbit 8:\\ 1440 =
5(288). \\
Orbit 9:\\ 2400 =
3(288)+
2(576)+
2(96)+
192. \\
Orbit 10:\\ 3600 =
4(288)+
4(576)+
144. \\
Orbit 11:\\ 7200 =
10(576)+
5(288). \\
Orbit 12:\\ 7200 =
5(288)+
10(576). \\
Orbit 13:\\ 7200 =
5(288)+
10(576). \\
Orbit 14:\\ 7200 =
10(576)+
5(288). \\
Orbit 15:\\ 14400 =
25(576). \\
\end{appendix}



\end{document}